\documentclass[12pt]{article}
\usepackage{amssymb,amsmath,epsfig}
\allowdisplaybreaks

\begin{document}
\title{\bf Study of Homogeneous and Isotropic Universe in $f(R,T^{\varphi})$ Gravity}
\author{M. Sharif \thanks{msharif.math@pu.edu.pk} and Aisha Siddiqa
\thanks{aisha.siddiqa17@yahoo.com}\\
Department of Mathematics, University of the Punjab,\\
Quaid-e-Azam Campus, Lahore-54590, Pakistan.}

\date{}
\maketitle

\begin{abstract}
This paper is devoted to study the cosmological behavior of
homogeneous and isotropic universe model in the context of
$f(R,T^{\varphi})$ gravity where $\varphi$ is the scalar field. For
this purpose, we follow the first order formalism defined by
$H=W(\varphi)$. We evaluate Hubble parameter, effective equation of
state parameter $(\omega^{eff})$, deceleration parameter and
potential of scalar field for three different values of
$W(\varphi)$. We obtain phantom era in some cases for the early
times. It is found that exponential expression of $W(\varphi)$
yields $\omega^{eff}$ independent of time for flat universe and
independent of model parameter otherwise. It is concluded that our
model corresponds to $\Lambda$CDM for both initial as well as late
times.
\end{abstract}

\section{Introduction}

The most striking and fascinating area in cosmology is the current
accelerated expansion of the universe suggested by various
observations. The agent causing this expansion is termed as dark
energy and it violates the strong energy condition. Cosmological
observations reveal that our universe is approximately homogeneous
and isotropic at large scales \cite{a} described by the standard FRW
model. Inclusion of cosmological constant ($\Lambda$) to the
standard model leads to $\Lambda$ cold dark matter ($\Lambda$CDM)
model. In general relativity (GR), the $\Lambda$CDM model explains
the expanding behavior of the universe where $\Lambda$ is supposed
to play the role of dark energy.

Despite its many features, there are two major issues associated
with this model named as fine tuning and coincidence problems
\cite{2}. The huge difference between the vacuum energy density and
the ground state energy suggested by quantum field theory leads to
the first issue. For this model, the densities of dark matter and
dark energy are of the same order leading to cosmological
coincidence problem. In order to resolve such and other open issues,
alternative models of dark energy were proposed either by modifying
matter or geometric part of the Einstein-Hilbert action. Modified
matter models \cite{3} are quintessence, phantom, K-essence,
holographic dark energy and Chaplygin gas. Some examples of
modification in geometric part are scalar-tensor theory, $f(R)$ and
$f(R,T)$ theories of gravity.

Harko \textit{et al}. \cite{7} proposed $f(R,T)$ gravity as a
generalized modified theory where $R$ is the Ricci scalar and $T$ is
the trace of energy-momentum tensor. The dependence on $T$ is
included due to the considerations of exotic fluids or quantum
effects. The coupling of curvature and matter yields interesting
consequences such as covariant derivative of the energy-momentum
tensor is no longer zero implying the existence of an extra force as
well as non-geodesic path of particles. In cosmological scenario, it
can explain the problem of galactic flat rotation curves as well as
dark matter and dark energy interactions \cite{7a}. Jamil \textit{et
al}. \cite{8} introduced some cosmic models in this gravity and
showed that dust fluid can reproduce $\Lambda$CDM model. Sharif and
Zubair \cite{10} explored thermodynamics and concluded that
generalized second law of thermodynamics is valid for phantom as
well as non-phantom phases. The same authors \cite{11} established
energy conditions and constraints for the stability of power-law
models.

Singh and Singh \cite{12a} discussed the reconstruction of $f(R,T)$
gravity models in the presence of perfect fluid and showed that
$f(R,T)$ extra terms can represent phantom dark energy as well as
cosmological constant in the presence and absence of perfect fluid,
respectively. Singh and Kumar \cite{12} explored the role of bulk
viscosity using FRW model with perfect fluid and found that bulk
viscosity provides a supplement for expansion. In literature
\cite{14}-\cite{16}, the higher dimensions are also explored in the
framework of $f(R,T)$ gravity. Moraes \textit{et al.} \cite{13}
studied hydrostatic equilibrium condition for neutron stars with a
specific form of equation of state (EoS) and found that the extreme
mass can cross observational limits.

In the paper \cite{7}, Harko \textit{et al.} also discussed
$f(R,T^{\varphi})$ theory where $T^{\varphi}$ is the trace of
energy-momentum tensor of a scalar field. Scalar fields have
extensively been studied in cosmology during the last three decades.
Yukawa is the pioneer to introduce scalar field in Physics which
naturally has an EoS $p=-\rho$ and hence is expected to play a vital
role in modern cosmology as well as astrophysics \cite{17}. Alves
\textit{et al.} \cite{17aa} confirmed the existence of gravitational
waves for extra polarization modes in both $f(R,T)$ as well as
$f(R,T^{\varphi})$ theories.

Halliwell \cite{18} explored the role of scalar field with
exponential potential and also discussed some examples that yield
exponential potential. Virbhadra \textit{et al.} \cite{19} studied
the effect of scalar field in gravitational lensing and explored the
features of lens specified by mass and charge of scalar field. Nunes
and Mimoso \cite{17a} worked on phase-plane analysis for a flat FRW
model in the presence of perfect fluid as well as self-interacting
scalar field. They showed that the scalar field potential having
positive, monotonic and asymptotically exponential behavior leads to
global attractor. Das and Banerjee \cite{17s} explored that the
scalar field can produce deceleration and acceleration phases of
cosmos by considering an energy transfer between scalar field and
dark matter.

Bazeia \textit{et al.} \cite{20} introduced the first order
formalism for scalar field models and also discussed some examples
of cosmological interest. In \cite{21}, the authors extended this
formalism for scalar fields with tachyonic dynamics. Recently,
Moraes and Santos \cite{23} presented a cosmological picture in
$f(R,T^{\varphi})$ gravity in the absence of matter by considering
flat FRW model. They started with the relation
$\dot{\varphi}=-\frac{dW}{d\varphi}$ which comes from the field
equations of GR via above mentioned formalism. We have applied this
formalism to Bianchi type-I universe model (with no matter present)
and concluded that the exponential form of $W(\varphi)$ can explain
all the stages of the universe evolution \cite{22}.

The motivation for neglecting matter action is that $f(R,T)$
framework cannot describe the radiation-dominated era because in
that phase the value of $T$ for perfect fluid is zero and $f(R,T)$
theory is reduced to $f(R)$ scenario \cite{23}. Moreover, Alves
\textit{et al.} \cite{17aa} showed that for gravitational wave
passing through vacuum with no extra polarization modes are induced
due to curvature matter coupling as $T=0$ in vacuum. On the other
hand, the same phenomenon in $f(R,T^{\varphi})$ is controlled by the
assumed model. Hence the study of radiation phase as well as any
phenomenon in vacuum has become a crucial issue to be addressed. In
this regard, Moraes introduced an extra dimension \cite{23s} and in
\cite{23ss} he considered a varying speed of light to obtain the
solutions in radiation phase.

In this paper, we explore cosmology of generalized FRW universe
model using the same action used in \cite{23} and with the
fundamental assumption of first order formalism, i.e.,
$H=W(\varphi)$ in $f(R,T^{\varphi})$ gravity. The plan of the paper
is as follows. In the next section, we formulate the field equations
and find the values of $H$, $\omega^{eff}$, $q$ and $V(\varphi)$. We
also discuss graphical behavior of these parameters for different
models of $W(\varphi)$. The last section provides the obtained
results.

\section{Field Equations and First Order Formalism}

The action for $f(R,T^{\varphi})$ gravity is given by
\begin{equation}\label{1}
S=\int d^{4}x\sqrt{-g}\left[{f(R, T^{\varphi})+\mathcal{L}(\varphi,
\partial_{\nu}\varphi)}\right],
\end{equation}
where we assume that $16\pi G=c=1$. In the following, we use the
model $f(R, T^{\varphi})=-\frac{R}{4}+\lambda T^{\varphi}$
\cite{23}, where $\lambda$ is a constant known as model parameter.
The corresponding field equations are
\begin{equation}\label{2}
G_{\alpha\beta}=2(T_{\alpha\beta}^{\varphi}-g_{\alpha\beta}\lambda
T^{\varphi} -2\lambda\partial_{\alpha}\varphi\partial_{\beta}
\varphi),
\end{equation}
where $G_{\alpha\beta}$ is the Einstein tensor and
$T_{\alpha\beta}^{\varphi}$ represents the energy-momentum tensor of
a scalar field. For a real $\varphi$, the Lagrangian density and the
energy-momentum tensor are given by
\begin{eqnarray}\label{3}
\mathcal{L}&=&\frac{1}{2}\partial_{\alpha}\varphi\partial^{\alpha}\varphi-V(\varphi),\\\label{4}
T_{\alpha\beta}^{\varphi}&=&\partial_{\alpha}\varphi\partial_{\beta}\varphi-
g_{\alpha\beta}\mathcal{L},
\end{eqnarray}
where $V(\varphi)$ denotes the self-interacting potential. The trace
of the energy-momentum tensor is
\begin{equation}\label{5}
T^{\varphi}=\dot{\varphi}^{2}+4V(\varphi),
\end{equation}
dot denotes derivative with respect to $t$.

The line element of FRW model is given by
\begin{equation}\label{6}
ds^{2}=-dt^{2}+a^{2}(t)(\frac{dr^{2}}{1-kr^{2}}+r^{2}(d\theta^{2}+\sin^{2}\theta
d\phi^{2})),
\end{equation}
where $a(t)$ is the scale factor and $k$ represents curvature of the
space. For $k=0,~1,~-1$, we have flat, closed and open universe
model, respectively. The corresponding field equations are
\begin{eqnarray}\label{7}
\frac{3}{2}H^{2}&=&(\frac{1}{2}-\lambda)\dot{\varphi}^{2}+
(4\lambda-1)V-\frac{3}{2}\frac{k}{a^{2}},\\\label{8}
\frac{3}{2}H^{2}+\dot{H}&=&-(\frac{1}{2}-\lambda)\dot{\varphi}^{2}+
(4\lambda-1)V-\frac{1}{2}\frac{k}{a^{2}},
\end{eqnarray}
which yield the expression of $\dot{H}$ as
\begin{equation}\label{a}
\dot{H}=-(1-2\lambda)\dot{\varphi}^{2}+\frac{k}{a^{2}}.
\end{equation}
The equation of motion for scalar field is obtained as
\begin{equation}\label{bb}
(1-2\lambda)(\ddot{\varphi}+3H\dot{\varphi})+(1-4\lambda)V_{\varphi}=0,
\end{equation}
where subscript $\varphi$ indicates derivative with respect to
$\varphi$. Following the first order formalism \cite{20}, the Hubble
parameter is given by
\begin{equation}\label{9}
H=W(\varphi).
\end{equation}
The expressions of $V(\varphi)$ and $\omega^{eff}$ from the field
equations become
\begin{eqnarray}\label{11}
V(\varphi)&=&\frac{1}{4\lambda-1}\left[\frac{3}{2}W^{2}-(\frac{1}{2}-\lambda)\dot{\varphi}^{2}+
\frac{3}{2}\frac{k}{a^{2}}\right], \\\label{10}
\omega^{eff}&=&\frac{p^{eff}}{\rho^{eff}}=-\left[1+\frac{\dot{\varphi}^{2}(4\lambda-2)}{3(W^{2}+
\frac{k}{a^{2}})}\right],
\end{eqnarray}
where $\rho^{eff}$ and $p^{eff}$ are
\begin{eqnarray}\nonumber
\rho^{eff}&=&(\frac{1}{2}-\lambda)\dot{\varphi}^{2}+
(1-4\lambda)V,\\\nonumber
p^{eff}&=&(\frac{1}{2}-\lambda)\dot{\varphi}^{2}-(1-4\lambda)V,
\end{eqnarray}
while the decelerating parameter is defined as
\begin{equation}\label{10a}
q=\frac{1}{2}(1+3\omega^{eff})\left(1+\frac{1}{a^{2}H^{2}}\right).
\end{equation}
Substituting $H$ from Eq.(\ref{9}) into (\ref{a}), we have
\begin{equation}\label{b}
(1-2\lambda)\dot{\varphi}^{2}+W_{\varphi}\dot{\varphi}-\frac{k}{a^{2}}=0,
\end{equation}
which is a quadratic equation in $\dot{\varphi}$. It has two roots
\begin{equation}\label{b1}
\dot{\varphi}=\frac{-W_{\varphi}\pm\sqrt{W_{\varphi}^{2}+\frac{4k}{a^{2}}(1-2\lambda)}}{2(1-2\lambda)},
\end{equation}
each of which is a first order differential equation. We find the
solution for the following three expressions of $W(\varphi)$.
\begin{enumerate}
\item $W(\varphi)=e^{b_{1}\varphi}$, where $b_{1}$ is a real constant. The
scalar field potentials followed by this value of $W(\varphi)$ are
of much interest. When $b_{1}=1$, it reproduces some negative
potential and $b_{1}=2$ leads to potential which presents
spontaneous symmetry breaking \cite{20}.
\item $W(\varphi)=b_{2}(\frac{\varphi^{2}}{3}-\varphi)$, where $b_{2}$
is a real constant. It represents a $\varphi^{4}$ type of model
\cite{27a}.
\item  $W(\varphi)=b_{3}\sin\varphi$, $b_{3}$ is a real constant.
It is a sine-Gordon type of model \cite{27b}.
\end{enumerate}

\subsection{Flat Universe $(k=0)$}

In this case, Eq.(\ref{b1}) implies that either
\begin{equation}\label{12}
\dot{\varphi}=0 \quad \text{or} \quad
\dot{\varphi}=-\frac{W_{\varphi}}{1-2\lambda}.
\end{equation}
For $\dot{\varphi}=0$, we have constant value of $\varphi$ and
consequently Eq.(\ref{9})-(\ref{10a}) yield
\begin{eqnarray}\nonumber
H=\text{constant}, \quad \omega^{eff}=-1, \quad q=-1,\quad
V(\varphi)=\text{constant},
\end{eqnarray}
which represents $\Lambda$CDM model. The solution of the second
option in Eq.(\ref{12}) for all forms of $W(\varphi)$ are given in
Table \textbf{1}, where $c_{1},~c_{2}$ and $c_{3}$ are constants of
integration. The graphical behavior of the corresponding $H$,
$\omega^{eff}$, $q$ and $V(\varphi)$ is shown in Figures
\textbf{1-3}. For graphical analysis, we have taken the free
parameters such that when $\omega^{eff}<-1$, $H$ is increasing while
if $\omega^{eff}>-1$, then $H$ is decreasing and when
$\omega^{eff}=-1$, $H$ is constant \cite{33}. According to standard
cosmology \cite{34}, $H\propto t^{-1}$ after inflation and it
becomes constant with the passage of time. We see that the Hubble
parameter shows this behavior only for exponential value of
$W(\varphi)$. The Hubble parameter and $\lambda$ are inversely
related, i.e., decrease in $\lambda$ increases $H$ and vice-versa
for all values of $W(\varphi)$.\\\\
Table \textbf{1}: Solution of
$\dot{\varphi}=-\frac{W_{\varphi}}{1-2\lambda}$
\begin{table}[bht]
\centering
\begin{small}
\begin{tabular}{|c|c|}
\hline\textbf{$W(\varphi)$}&\textbf{$\varphi(t)$}\\
\hline$e^{b_{1}\varphi}$&$\frac{1}{b_{1}}\ln\left[\frac{1-2\lambda}{b_{1}^{2}(t+c_{1})}\right]$\\
\hline$b_{2}(\frac{\varphi^{3}}{3}-\varphi)$&$\tanh\left[\frac{b_{2}(t+c_{2})}{1-2\lambda}\right]$\\
\hline$b_{3}\sin\varphi$&$2\arctan\left[\tanh\left[\frac{b_{3}t}{4\lambda-2}+\frac{c_{3}}{2}\right]\right]$\\
\hline
\end{tabular}
\end{small}
\end{table}
\begin{figure}
\epsfig{file=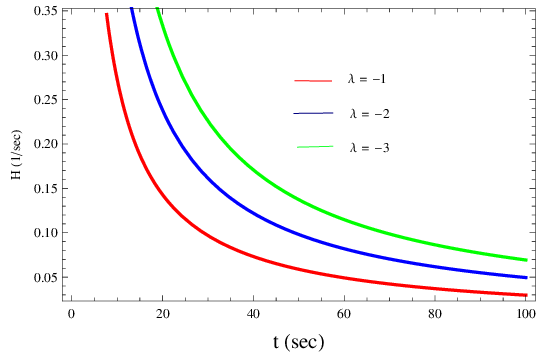,width=0.5\linewidth}
\epsfig{file=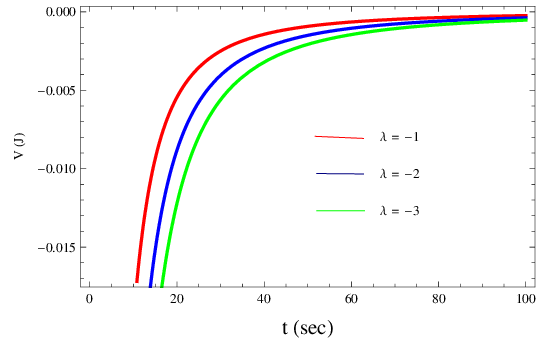,width=0.5\linewidth} \caption{Plots of $H$ and
$V$ versus $t$ for
$W(\varphi)=e^{b_{1}\varphi},~b_{1}=1(1/sec),~c_{1}=1$ when $k=0$.}
\end{figure}
\begin{figure}
\epsfig{file=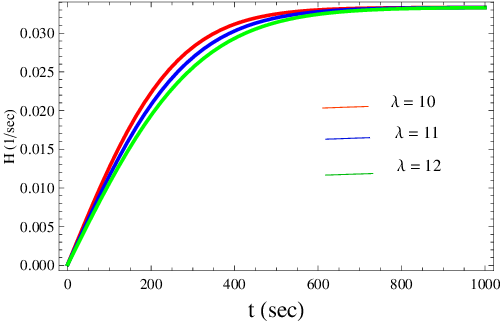,width=0.5\linewidth}
\epsfig{file=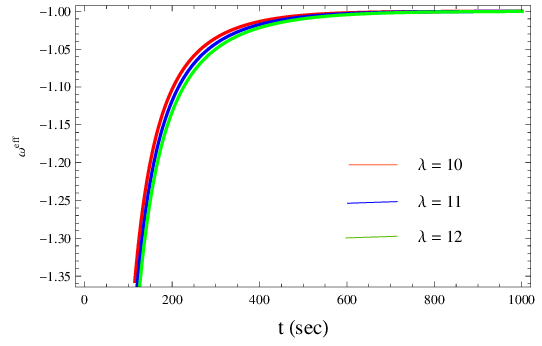,width=0.5\linewidth}
\epsfig{file=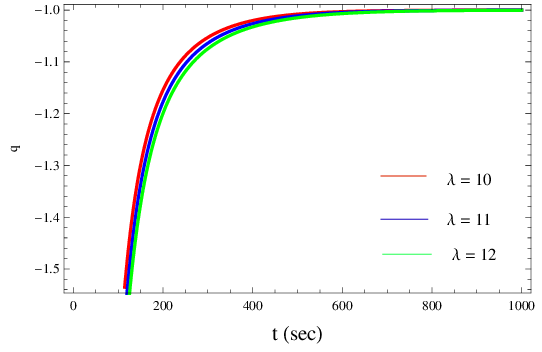,width=0.5\linewidth}
\epsfig{file=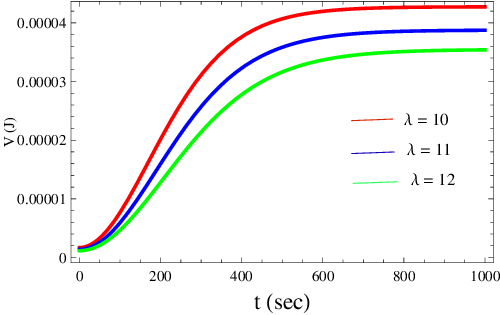,width=0.5\linewidth} \caption{Plots of $H$,
$\omega^{eff}$, $q$ and $V$ versus $t$ for
$W(\varphi)=b_{2}(\frac{\varphi^{2}}{3}-\varphi),~b_{2}=0.05(1/sec),~c_{2}=1$
 when $k=0$.}
\end{figure}
\begin{figure}
\epsfig{file=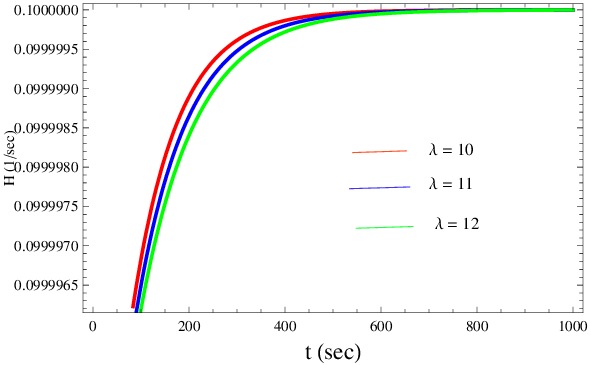,width=0.5\linewidth}
\epsfig{file=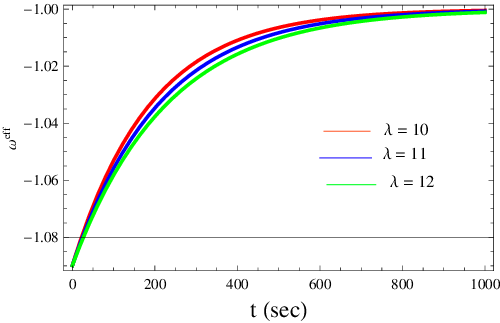,width=0.5\linewidth}
\epsfig{file=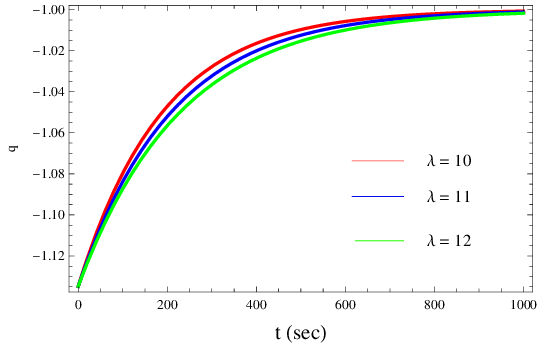,width=0.5\linewidth}
\epsfig{file=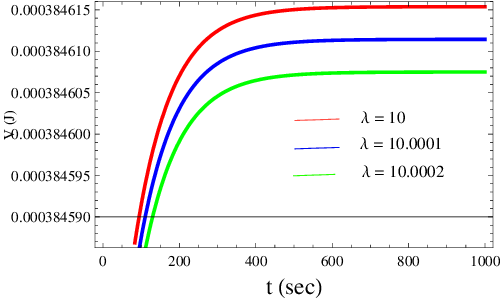,width=0.5\linewidth} \caption{Plots of $H$,
$\omega^{eff}$, $q$ and $V$ versus $t$ for
$W(\varphi)=b_{3}\sin\varphi,~b_{3}=0.1(1/sec),~c_{3}=5$ when
$k=0$.}
\end{figure}

The exponential value of $W(\varphi)$ yields the following time
independent expressions for $\omega^{eff}$ and $q$
\begin{eqnarray}\label{10b}
\omega^{eff}&=&-1+\frac{2b_{1}^{2}}{3(1-2\lambda)},\\\label{10c}
q&=&-1+\frac{b_{1}^{2}}{(1-2\lambda)},
\end{eqnarray}
Eq.(\ref{10b}) shows that $\omega^{eff}$ can have different values
depending upon the arbitrary constants $b_{1}$ and $\lambda$. For
polynomial and trigonometric forms of $W(\varphi)$, it initially
falls in phantom regime and then approaches to $\Lambda$CDM model as
time increases. The increase in $\lambda$, decreases $\omega^{eff}$
for the second and third forms of $W(\varphi)$. We see that
$\omega^{eff}$ is $-1$ for all values of $\lambda$ at late times.
The value of deceleration parameter, given in Eq.(\ref{10c})
indicates that we have accelerated expansion phase when
$b_{1}^{2}<1-2\lambda$ and vice versa. The graphs of $q$ in Figures
\textbf{2} and \textbf{3} show that the rate of expansion is
decreasing with time. The potential of scalar field is negative as
well as increasing for the first form while positive and increasing
for the remaining values of $W(\varphi)$. Also, it decreases for the
first value and increases for the other two with decrease in
$\lambda$.

\subsection{Closed Universe $(k=1)$}

\begin{figure}
\epsfig{file=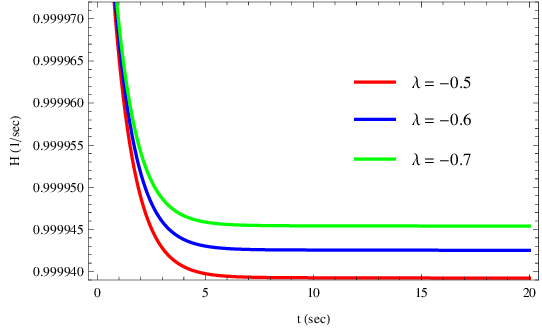,width=0.5\linewidth}
\epsfig{file=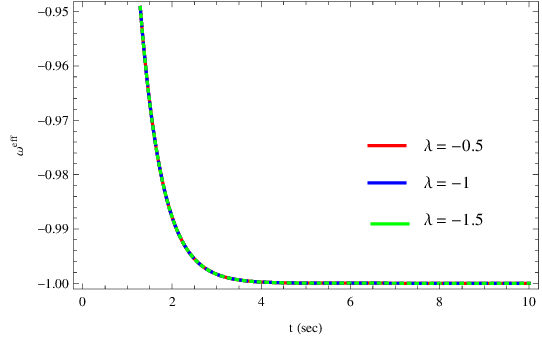,width=0.5\linewidth}
\epsfig{file=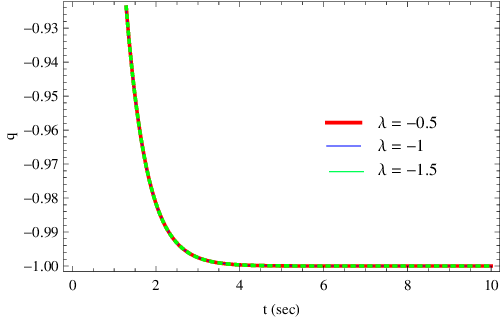,width=0.5\linewidth}
\epsfig{file=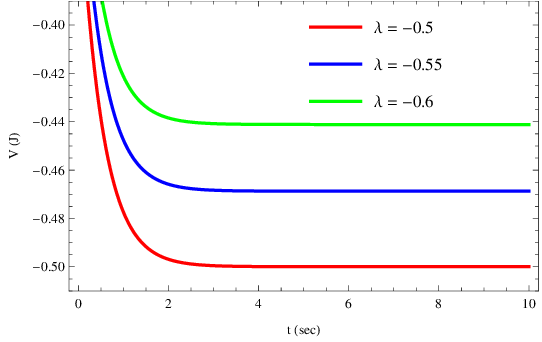,width=0.5\linewidth} \caption{Plots of $H$,
$\omega^{eff}$, $q$ and $V$ versus $t$ for
$W(\varphi^{-})=e^{b_{1}\varphi^{-}},~b_{1}=0.0001(1/sec),~
\varphi^{-}(0)=0.1$ when $k=1$.}
\end{figure}
\begin{figure}
\epsfig{file=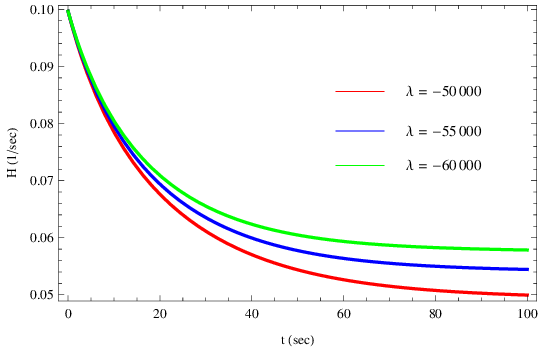,width=0.5\linewidth}
\epsfig{file=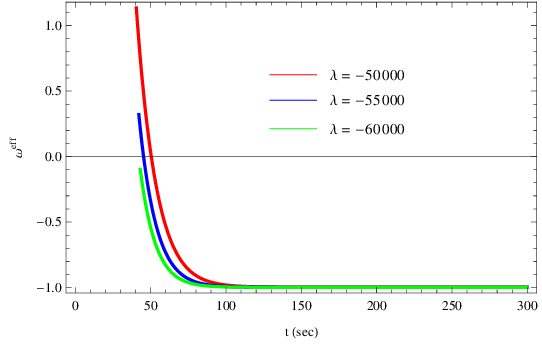,width=0.5\linewidth}
\epsfig{file=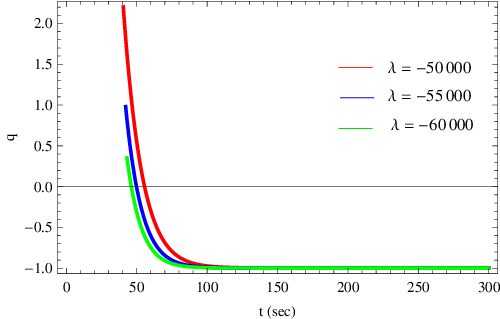,width=0.5\linewidth}
\epsfig{file=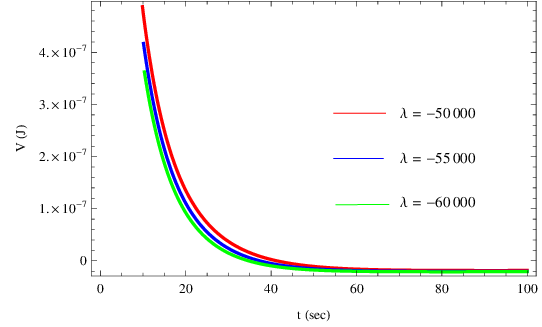,width=0.5\linewidth} \caption{Plots of $H$,
$\omega^{eff}$, $q$ and $V$ versus $t$ for
$W(\varphi^{-})=b_{2}(\frac{(\varphi^{-})^{2}}{3}-\varphi^{-}),~b_{2}=-1(1/sec),~\varphi^{-}(0)=0.1$
when $k=1$.}
\end{figure}
\begin{figure}
\epsfig{file=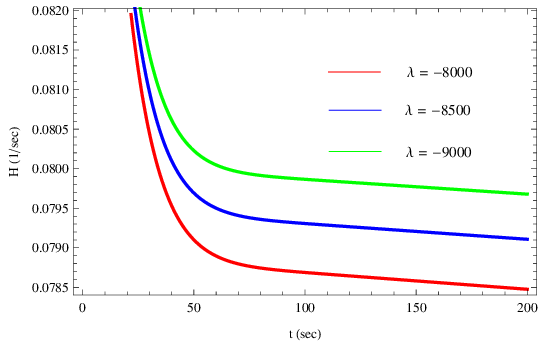,width=0.5\linewidth}
\epsfig{file=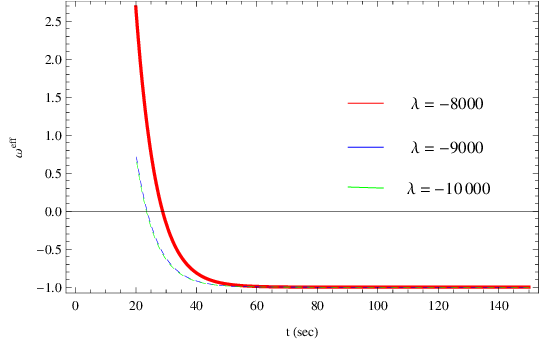,width=0.5\linewidth}
\epsfig{file=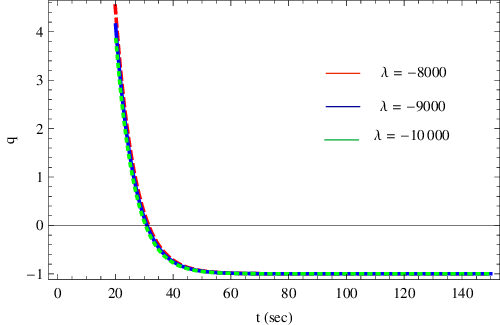,width=0.5\linewidth}
\epsfig{file=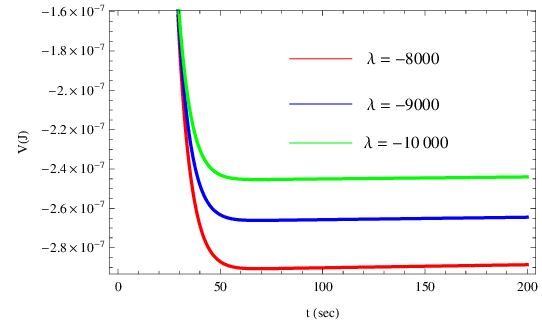,width=0.5\linewidth} \caption{Plots of $H$,
$\omega^{eff}$, $q$ and $V$ versus $t$ for
$W(\varphi^{-})=b_{3}\sin\varphi^{-},~b_{3}=0.2(1/sec),~\varphi^{-}(0)=0.5$
when $k=1$.}
\end{figure}

In this case, Eq.(\ref{b1}) gives two roots $\varphi^{-}$ and
$\varphi^{+}$ which are solved numerically for the three forms of
$W(\varphi)$. The plots of $H,~\omega^{eff},~q$ and $V(\varphi)$ for
$\varphi^{-}$ are shown in Figures \textbf{4-6}. The plots of $H$
show standard behavior for all forms of $W(\varphi)$ while decrease
in $\lambda$ increases $H$. The plots of $\omega^{eff}$ indicates
that for all $W(\varphi)$, it is $-1$ or corresponds to $\Lambda$CDM
as time increases. The change in $\lambda$ does not affect
$\omega^{eff}$ for exponential form of $W(\varphi)$ while it
initially decreases with decrease in $\lambda$ for polynomial form
and increases with decrease in $\lambda$ for trigonometric case but
it is equal to $-1$ for all values of $\lambda$ as time increases.

The graphs of $q$ in this case indicate an accelerated expansion
which approaches to de-sitter expansion for late times. Also,
increase in $\lambda$ increases $q$ for second and third models of
scalar field while for first model $q$ remains constant for all
values of $\lambda$. The plots for scalar field potential show that
it is negative and decreasing for exponential as well as
trigonometric forms while positive and decreasing for the polynomial
expression. The decrease in $\lambda$ implies an increase in
$V(\varphi)$ for first and third values of $W(\varphi)$ but it has
an opposite effect for second value. The behavior of $H$,
$\omega^{eff}$ and $V(\varphi)$ for $\varphi^{+}$ is similar to
$\varphi^{-}$. For $\varphi^{+}$, the initial condition and model
parameter remain unchanged while the sign of $b_{i}~(i=1,2,3)$ is
opposite.

\subsection{Open Universe $(k=-1)$}

In this case, the graphs are given in Figures \textbf{7-9}. The
Hubble parameter has increasing behavior which approaches to
constant when the model approaches to $\Lambda$CDM for all the three
forms of $W(\varphi)$. For all forms, $\omega^{eff}$ initially lies
in phantom era and for later times it approaches to $-1$. The
behavior of $q$ indicates that the expansion rate is slowing down.
The scalar field potential is positive and decreasing for
exponential case while positive and increasing for the remaining
forms. The effect of model parameter is also observed in each graph.
All plots for $\varphi^{+}$ have the same behavior as for
$\varphi^{-}$ under the conditions defined in the previous case.
\begin{figure}
\epsfig{file=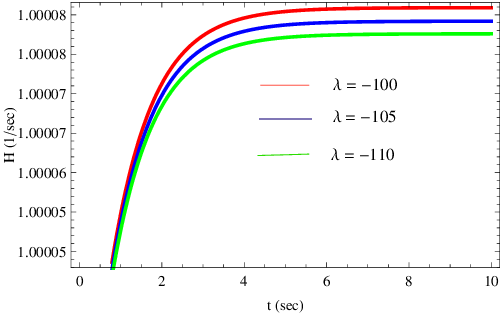,width=0.5\linewidth}
\epsfig{file=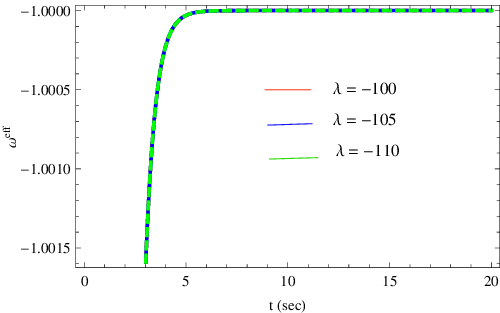,width=0.5\linewidth}
\epsfig{file=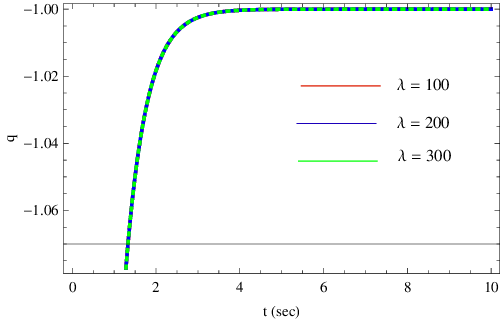,width=0.5\linewidth}
\epsfig{file=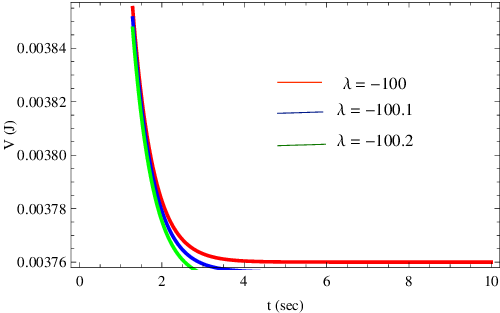,width=0.5\linewidth} \caption{Plots of $H$,
$\omega^{eff}$, $q$ and $V$ versus $t$ for
$W(\varphi^{-})=e^{b_{1}\varphi^{-}},~b_{1}=0.001(1/sec),~
\varphi^{-}(0)=0.01$ when $k=-1$.}
\end{figure}
\begin{figure}
\epsfig{file=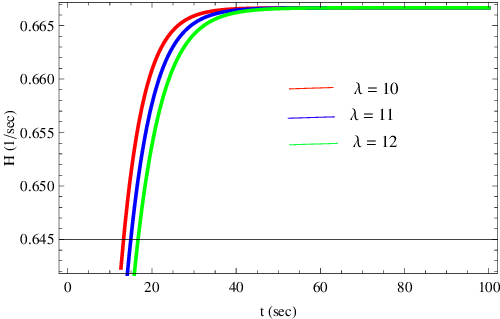,width=0.5\linewidth}
\epsfig{file=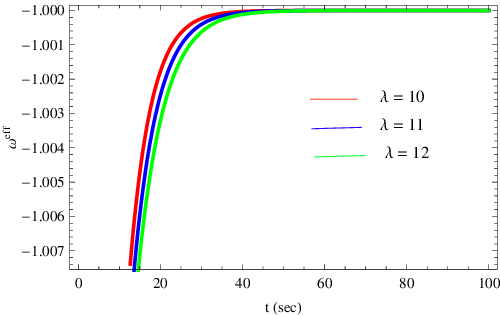,width=0.5\linewidth}
\epsfig{file=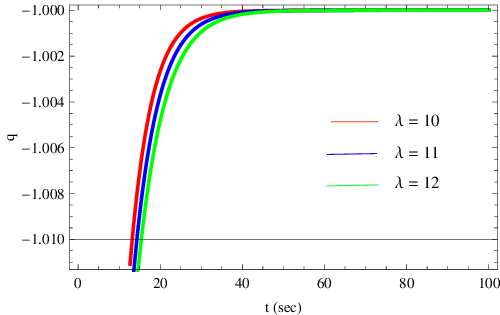,width=0.5\linewidth}
\epsfig{file=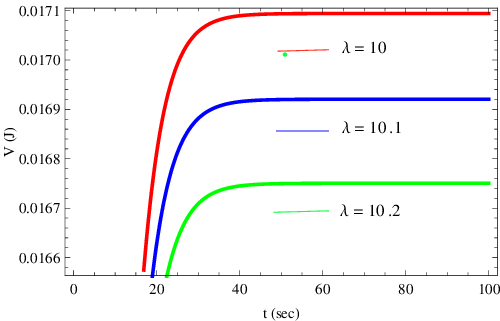,width=0.5\linewidth} \caption{Plots of $H$,
$\omega^{eff}$, $q$ and $V$ versus $t$ for
$W(\varphi^{-})=b_{2}(\frac{(\varphi^{-})^{2}}{3}-\varphi^{-}),~b_{2}=-1(1/sec),~\varphi^{-}(0)=0.1$
when $k=-1$.}
\end{figure}
\begin{figure}
\epsfig{file=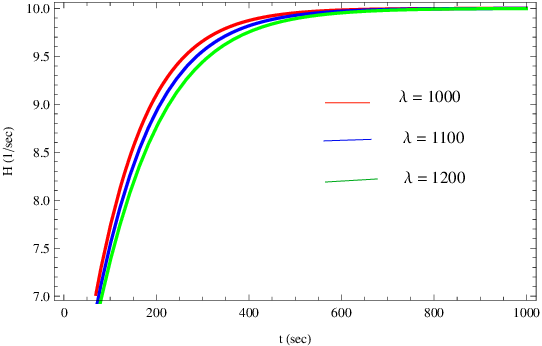,width=0.5\linewidth}
\epsfig{file=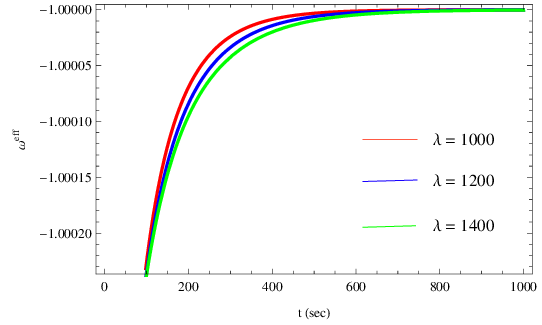,width=0.5\linewidth}
\epsfig{file=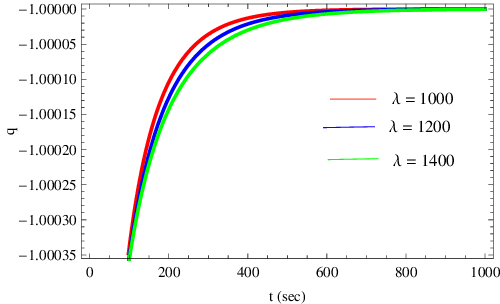,width=0.5\linewidth}
\epsfig{file=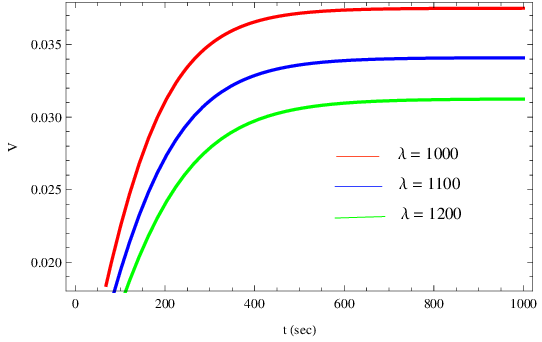,width=0.5\linewidth} \caption{Plots of $H$,
$\omega^{eff}$, $q$ and $V$ versus $t$ for
$W(\varphi^{-})=b_{3}\sin\varphi^{-},~b_{3}=10(1/sec),~\varphi^{-}(0)=0.5$
when $k=-1$.}
\end{figure}

\section{Concluding Remarks}

This paper investigates the cosmological behavior of FRW universe
model in the background of $f(R,T^{\varphi})$ gravity and in the
absence of matter. We studied flat $(k=0)$, spherical $(k=1)$ and
hyperbolic $(k=-1)$ universe models assuming different values of
Hubble parameter as a function of scalar field $\varphi$. It is
found that in flat universe, our model corresponds to $\Lambda$CDM
for exponential form of Hubble parameter while for other two forms,
it represents phantom phase in the beginning and then $\Lambda$CDM
for late times. For the closed universe, quintessence phase is shown
at early times and $\Lambda$CDM model for late times for all forms
of $W(\varphi)$. In closed universe the polynomial and trigonometric
expressions of $W(\varphi)$ described the stages of stiff fluid,
radiation-dominated phase, matter-dominated era, quintessence and
de-Sitter expansion. In open universe scenario, our model initially
falls in phantom phase and then is analogous to $\Lambda$CDM for all
values of $W(\varphi)$. This is the main significance of our model
that it can characterize the standard model for early as well as
late time.

The deceleration parameter depicts the rate of expansion is growing
for closed universe while it is slowing down for open universe. In
case of flat universe, it is constant for exponential value and for
remaining expressions of $W(\varphi)$, $q$ has the same behavior as
for open universe. The potential of scalar field has positive as
well as negative values for different cases. It is important to note
that when $V(\varphi)$ is increasing negatively (Figure \textbf{4}),
the corresponding $\omega^{eff}$ approaches to $-1$ much earlier
than other cases. The graphical analysis indicates different effects
of the model parameter for different cases. It is interesting to
mention here that $\omega^{eff}$ and $q$ remain the same with
varying $\lambda$ for $k=\pm1$ when $W(\varphi)$ is in exponential
form. In view of the above discussion, we conclude that the closed
universe scenario provides consistent results with an expanding
universe in the context of $f(R,T^{\varphi})$ gravity.

Finally, we compare our results for the flat universe model with the
paper \cite{23}, in which the authors used the differential equation
$\dot{\varphi}=-W_{\varphi}$ obtained from GR field equations
\cite{20}. They found that the model
$f(R,T^{\varphi})=-\frac{R}{4}+\lambda T^{\varphi}$ yields
$-1<\omega^{eff}<\frac{1}{3}$ which does not correspond to the
phantom era. However, we have used the first order formalism in its
true spirit, i.e., assuming Hubble parameter as a function of scalar
field and use modified field equations to obtain the differential
equation Eq.(\ref{b1}). We have found that this model characterizes
$\Lambda$CDM for exponential $H$ and represents $\Lambda$CDM as well
as phantom phase for other two forms of $W(\varphi)$.

\vspace{0.50cm}

{\bf Acknowledgment}

\vspace{0.25cm}

We would like to thank the Higher Education Commission, Islamabad,
Pakistan for its financial support through the {\it Indigenous Ph.D.
5000 Fellowship Program Phase-II, Batch-III}.

\end{document}